\documentclass[prb,showpacs,twocolumn]{revtex4}

\usepackage[dvips]{graphicx}
\usepackage{dcolumn}
\usepackage{bm}
\def\be{\begin{equation}}
\def\en{\end{equation}}

\def\p{\partial}

\def\ls{\lesssim}
\newcommand{\bi}[1]{\mbox{\boldmath$#1$}}
\def\p{\partial}
\def\bea{\begin{eqnarray}}
\def\ena{\end{eqnarray}}

\renewcommand{\theequation}{\arabic{section}.\arabic{equation}}

\def\ve{\varepsilon}

\begin{document}

\title{Surface tension of electrolytes: 
Hydrophilic and hydrophobic ions near an interface}

\author{Akira Onuki}
\affiliation{Department of Physics, Kyoto University, Kyoto 606-8502,
Japan}
\date{\today}

\begin{abstract}
We calculate the ion distributions 
around an interface in fluid mixtures of highly polar 
and less polar fluids (water and oil)  for two and three 
ion species. We take into account the solvation 
and image interactions between ions and solvent. 
We show that hydrophilic and hydrophobic ions 
tend to undergo a microphase separation at an 
interface,  giving rise  to   an enlarged   electric 
double layer. We also derive 
a general expression for the surface tension 
of electrolyte systems, which   contains a  negative electrostatic 
contribution  proportional to the square root 
of the bulk salt density.  The amplitude of this square-root term 
is small for hydrophilic ion pairs, but is much increased 
for  hydrophilic and hydrophobic ion pairs. 
For three ion species including hydrophilic and hydrophobic 
ions, we  calculate 
the ion distributions to explain  those 
obtained by   x-ray reflectivity measurements.  
\end{abstract}


\maketitle


\section{Introduction}

It has  long been known    that 
the surface tension $\gamma$ of 
a water-air interface  increases with 
addition of   inorganic salts    in water. 
Wagner \cite{Wagner}  ascribed its origin to 
the image charge in air due to the difference 
in the dielectric constants of  the two phases,  
which repels each ion in water  away from the interface. 
Onsager and Samaras  \cite{Onsager} obtained the 
limiting law for the surface tension change 
in the form 
$\Delta\gamma= \frac{1}{8} T n_{\rm w}\ell_{\rm B} 
[\ln (1/n_{\rm w}\ell_{\rm B}^3)+$const.$]$,
where $n_{\rm w} (\ll \ell_{\rm B}^{-3})$ is the ion  density  
in the bulk water and $\ell_{\rm B}$ is the Bjerrum 
length.  Levin and Flores-Mena \cite{Levin} 
took account of ion depletion due to a relatively large size of 
the hydration  shell radius 
\cite{Is}.  In experiments, 
at not extreme  dilution, 
the linear behavior 
$\Delta\gamma=T n_{\rm w}\lambda_s$ 
 has been  measured  
\cite{Jones,tension1,Ohshima,Petersen},    
where  $\lambda_s$ is the effective thickness 
of the ion-free layer. For example, 
 $\lambda_s \sim 3{\rm \AA}$ 
for NaCl, where the densities of Na$^+$ and Cl$^-$ are 
 $n_{\rm w}/2$. However, for a number of salts around 1 mM 
   in aqueous solutions, 
Jones and Ray \cite{Jones} detected a very small negative minimum 
in $\Delta\gamma$, which still remains an unsolved  controversy 
\cite{Petersen,Ka,Manciu}.

In polar fluid mixtures, the ion distributions 
are  much more complicated 
 when the ions  are soluble in the two phases 
 \cite{Nitro,Onuki-Kitamura,OnukiPRE}.   
In this problem we need to 
account for the  solvation 
between ions and solvent molecules 
(hydration in aqueous solutions), whose 
  free energy contribution  usually 
 much exceeds  the thermal energy 
$T$ (per ion)  \cite{Is}. 
In particular, 
 one ion species can prefer  
one  fluid component, 
while  the other species can prefer  
the other  component. Such asymmetric ion pairs 
tend to segregate around  an interface, while the 
segregation is prohibited on larger scales 
due to the charge neutrality in the bulk. 
In this paper, we will 
examine this microphase separation near an interface 
to calculate a  decrease of the 
surface tension $\Delta\gamma<0$, 
whose amplitude  can be 
much larger than that of the well-known 
increase for hydrophilic ion  pairs.  
It is  worth noting 
that, in their small-angle neutron 
scattering experiment,  Sadakane {\it et al.}  
\cite{Seto} found  periodic charge-density-wave 
 structures  in a near-critical binary mixture of 
D$_2$O-trimethylpyridine 
containing sodium tetrarphenylborate (NaBPh$_4$). 
Their  salt is composed of 
strongly hydrophilic cation Na$^+$   and strongly hydrophobic anion 
BPh${_4}^{-}$, 
which should  considerably  decrease the surface tension 
and produce a mesoscopic structure near 
the critical point.

In their   x-ray reflectivity experiment,   
Luo {\it et al.} \cite{Luo} 
 measured the ion distributions 
in the vicinity of an interface in water-nitrobenzene. 
They added two salts, tetrabutylammonium 
tetraphenylborate (TBA-TPB) 
and  tetrabutylammonium bromide (TBA-Br). 
Then they realized  a two-phase state   with    
hydrophilic anion Br$^-$, 
hydrophobic cation TBA$^+$,  
and hydrophobic anion TPB$^-$. 
They detected  strong accumulation or 
depletion of the ions 
on the two sides of  the interface, 
which suggest a crucial role of the 
ion-solvent interactions 
dependent on the ion and  solvent  species. 
In water-nitrobenzene containing both hydrophilic 
and hydrophobic ions, 
a large drop  of the surface tension 
has been observed \cite{Nitro,Luo}.

We mention the presence of 
a large body of spectroscopic experiments and 
computer simulations with molecular 
resolution on ion effects at  a water-air interface 
\cite{Benjamin,Tobias}, where the air region may be 
treated as a vacuum.   
Such studies provide detailed information 
of ionic interfacial  behavior on the angstrom scale for 
various ion species.   However, microscopic 
studies remain inadequate for ion effects 
at a water-oil interface. 
On the contrary, our approach in this 
paper  will be based on   a  
Ginzburg-Landau theory of solvation and ion distributions  
\cite{Onuki-Kitamura,OnukiPRE}. 
We will consider  water-oil systems like water-nitrobenzene,  
where the dielectric constants 
of the two components are not much separated 
(the dielectric constnt of nitrobenzene is about 35).

The organization of this paper is as folows.  
In Section 2, 
we will present a  Ginzburg-Landau approach to 
the molecular interactions between ions 
and solvent molecules. Taking account of   the electrostatic, 
solvation, and image interactions, we will 
introduce the grand potential and present a theoretical 
expression for  the surface tension. 
It contains a negative 
 electrostatic correction, which is not included in 
the Gibbs theory \cite{Gibbs,Widom} but is crucial 
for hydrophilic and hydrophobic ion pairs.    
In Section 3, we will numerically  examine the ion 
distributions and the surface tension 
for hydrophilic and hydrophobic  ion pairs. 
We will also discuss 
the behavior of $\Delta\gamma$ 
in the usual case of hydrophilic   
ion pairs in extreme dilution. 
In Section 4, we extend our theory in the presence 
 of three ion species 
to explain the experiment by Luo {\it et al.} \cite{Luo}. 
 In Appendix A, we will examine the relationship of 
our surface tension formula  and 
the Gibbs equation. 
In Appendix B, we will relate 
our grand potential to  the bulk pressure 
and derive the Laplace law using our surface tension formula, 
since we will treat incompressible 
binary mixtures in the text.

\section{Theoretical background}
\setcounter{equation}{0}
\subsection{Ginzburg-Landau free energy}

We consider a polar binary 
 mixture  containing a small amount of salt. 
The volume fraction  
of the more polar component  is  written as $\phi$. 
The other less polar  component has the volume fraction 
$1-\phi$.   We neglect the volume fractions of the ions. 
The ion  densities are 
written as  $n_1,n_2,\cdots$.   
Their charges are $Z_ie$, so 
 $Z_1=1$ and $Z_2=-1$  for two monovalent 
ion species.   In our scheme,  $\phi$, $n_1, n_2,\cdots$  
 are  smooth space-dependent 
  variables coarse-grained  
on the microscopic level. 
We examine the ion distribution around an 
interface  \cite{OnukiPRE}, 
where all the quantities change along the 
$z$ axis. The solvent molecular sizes are 
given by a common length $a$ for 
the two components.  Then $\phi$ also 
represents the molar concentration. 
The interface thickness $\xi$ is assumed  to be 
longer than  $a$. We neglect the  formation  
of  dipole pairs and ion clusters, which 
is relevant at not small  
ion densities \cite{pair,pair1,pair2}.

The free energy  $F$ of our system is the space integral 
of the free energy density $f$  of the form, 
\bea 
f&=& f_0(\phi,T) + \frac{C}{2}|\nabla \phi|^2+
\frac{\ve(\phi)}{8\pi}{ E}^2 \nonumber\\
&& \hspace{-1cm}+ 
T \sum_{i} {n_i} [\ln (n_ia^3) -g_i\phi] 
+  \mu_{\rm im}\sum_i Z_i^2 n_i  .
\ena   
We set the Boltzmann constant equal to unity. 
The first term $f_0$ is the chemical part in  
the Bragg-Williams form,  
\be 
f_0 = \frac{T}{a^3}[ 
 \phi \ln\phi + (1-\phi)\ln (1-\phi) 
+ \chi \phi (1-\phi) ], 
\en 
where $\chi$ depends on the temperature $T$ 
and its  critical value is 2 
in the absence of ions   \cite{Onukibook,Safran}. 
The  second term is the gradient part, 
while the third term is the 
electrostatic free energy with $\Phi$ 
being the electric potential. 
The electrostatic potential $\Phi$ satisfies   
$\nabla\cdot\ve(\phi)\nabla \Phi=- 
 4\pi \rho$, where 
  $\rho=\sum_i Z_ien_i$  is the charge density. 
Around an interface the electric field  
${ E}= -d\Phi/dz$  
 is expressed as  
\be 
E(z) = \frac{4\pi}{\ve(\phi(z))} 
\int_{-\infty}^zd z'\rho(z'),
\en  
where  the lower bound of the integration 
is pushed to $-\infty$. 
The dielectric constant $\ve$ 
is  assumed to be of  the linear form \cite{Debye-Kleboth}  
\be 
\ve(\phi)=\ve_c + \ve_1 (\phi-\phi_c) ,
\en 
where $\ve_c$ and $\ve_1$ are constants. 
 The $\phi_c$ is the critical volume fraction 
($=1/2$ for the free energy density in Eq.(2.2)).
Thus $\ve(\phi)$ depends on $z$ near an interface. 
The   coupling terms $-Tg_in_i\phi$ 
arise from the composition-dependence of 
the solvation chemical potentials \cite{OnukiPRE}. 
The differences  $Tg_i\Delta\phi$ 
 may be equated to  the Gibbs transfer 
energies (per particle here) known  
 in electrochemistry \cite{Nitro,Gros,Koryta,Hung,Osakai}, 
where $\Delta\phi$ is the concentration difference 
between the two phases. 
From data of room-temperature 
water-nitrobenzene in  strong segregation \cite{estimate},  
$g_i$   are expected to be typically of order 15 
for monovalent hydrophilic ions 
and are  even larger 
for multivalent ions such as Ca$^{2+}$ 
or  Al$^{3+}$.  For the hydrophobic 
anion BPh$_4^-$ (tetraphenylborate), for example,  
it is about $-15$, on the other hand. 
The resultant  solvation coupling 
between  the ions and the composition 
 is  thus very strong, dramatically affecting 
the phase transition behavior near the critical point 
 \cite{Onuki-Kitamura,OnukiPRE}. 
In binary mixtures in which the more polar 
component is water, the coupling constants 
$g_{i}$  are positive for hydrophilic  ions and  
negative for hydrophobic ions.

The last term in Eq.(2.1) 
represents the image interaction, 
arising  from 
inhomogeneous dielectric constant $\ve$. 
It originates from  the discrete nature of ions,   
while the electric field $\Phi$ in our theory  
is produced by the smoothly 
 coarse-grained charge density $\rho$. 
The original papers \cite{Wagner,Onsager} 
treated  water-air interfaces, 
but we suppose weak or moderate inhomogeneity 
of the dielectric constant across a diffuse interface.   
Then  it follows  the  Cauchy  integral form \cite{OnukiPRE},  
\be 
\mu_{\rm im}(z)= T Aa  
\frac{\ve_1}{\ve_c} 
\int  \frac{dz'}{\pi} 
\frac{{e^{-2\kappa |z-z'|}} }{z-z'} \frac{d\phi(z')}{dz'} ,
\en 
to first order in $\ve_1$. 
The coefficient  $A$ represents the charge 
strength as 
\be 
A= \pi e^2/4a\ve_c T=\pi \ell_{{\rm B}c}/4a
\en 
where $ 
\ell_{{\rm B}c}=e^2/\ve_c T$ 
is the Bjerrum length at $\ve=\ve_c$. 
The damping  factor $e^{-2\kappa |z-z'|}$ in Eq.(2.5) 
arises from the screening of the potential 
by the other ions \cite{Wagner,Onsager}.  
In our numerical analysis, we  treat  $\kappa$ 
 as the space-dependent local 
 value $[4\pi e^2m({\bi r})/\ve({\bi r})  T]^{1/2}$ 
with $m=\sum_iZ_i^2n_i$.

In equilibrium  we assume 
homogeneity of the chemical potentials 
$\mu_i= \delta F/\delta n_i$ ($i=1,2,\cdots)$  
and $h=\delta F/\delta \phi$. 
We introduce  normalized ion densities \cite{cc},  
\be 
c_i=a^3n_i    \quad (i=1,2,\cdots) . 
\en 
In our theory  the ion volume fractions 
are assumed to be so small such that 
the real ion sizes do not come into play. 
Since we have 
\be 
\mu_i= T(\ln c_i+1-g_i\phi)+Z_i^2 \mu_{\rm im} +Z_ie\Phi, 
\en
the ion density profiles are   expressed as 
\be
\frac{c_i(z)}{ c_{i0}} =  \exp
\bigg[g_i\phi(z) -\frac{Z_i{e}}{T} \Phi(z)
- \frac{Z_i^2}{T}\mu_{\rm im}(z) \bigg],
\en  
where  $c_{i0}=\exp(\mu_i/T-1)$. 
In taking the derivatives of the image free energy  
$F_{\rm im}=\int d{\bi r}\mu_{\rm im}\sum_i Z_i^2 n_i$  
 with respect to $n_i$,  
we neglect the $n$-dependence of $\mu_{\rm im}$ in Eq.(2.5). 
The  $h=\delta F/\delta\phi$ has the meaning of the chemical 
potential difference of the two fluid components 
(divided by $a^3$). It  
 is of the form,     
\be 
h=f_0'(\phi)
 -C\phi''  - \frac{\ve_1}{8\pi}{ E}^2  -T \sum_i g_in_i 
  + h_{\rm im},
\en 
where  $f_0'= {\p f_0}/{\p \phi}$ and  
$\phi'' = d^2\phi/d z^2$. 
From the image interaction 
we have the contribution  $h_{\rm im}= 
\delta F_{\rm im}/\delta \phi$. 
In the 1D case,   $h_{\rm im}$ is given by 
the right hand side of Eq.(2.5) with  
$d\psi(z')/dz'$ being  replaced by 
$\sum_i Z_i^2 dn_i(z')/dz'$.

Hereafter the quantities with the subscript $\alpha$ 
(the subscript $\beta$) denote 
the bulk values in the more (less) polar phase  
attained as  $z \rightarrow -\infty$ 
(as  $z \rightarrow \infty$). 
We assume that the screening lengths, $\kappa_\alpha^{-1}$ 
and  $\kappa_\beta^{-1}$,  are much shorter than the system length $L$.  
In real systems 
 this might not be the case for extremely  small $n_\beta$. 
From Eq.(2.9) the distribution ratio of each ion 
can be expressed as 
\be 
c_{i\beta}/c_{i\alpha} = \exp[{Z_ie\Delta\Phi/T-g_i\Delta\phi}], 
\en 
where  $\Delta\phi=\phi_\alpha-\phi_\beta$.  
The image interaction 
vanishes far from the interface and does not appear  
in Eq.(2.11).  There arises a Galvani potential difference 
$\Delta\Phi=\Phi_\alpha-\Phi_\beta$ across 
an interface. It is determined by the charge neutrality 
far from the interface, 
\be 
\sum_iZ_i c_{i\alpha}= 
\sum_iZ_i c_{i\beta}=0.
\en  
In addition,  as $z\rightarrow \pm \infty$, 
the homogeneity of $h$ in Eq.(2.10) yields the bulk relations,  
\bea 
h &=& f_0'(\phi_\alpha)   -T \sum_i g_in_{i\alpha}\nonumber\\ 
&=&f_0'(\phi_\beta)   -T \sum_i g_in_{i\beta}.
\ena  

\subsection{Image interaction}

We give more discussions on the image interaction. 
From Eq.(2.9) it is important  
under the condition 
\cite{OnukiPRE},
\be 
\xi< Z_i^2\ell_{{\rm B}c}\ve_1 \Delta\phi/4\ve_c < \kappa^{-1},
\en 
where $\kappa=\kappa_\alpha$ or $\kappa_\beta$ 
in the two phases.  When ions are present only in 
the more polar phase under Eq.(2.14), 
the change of the surface tension 
 obeys the  Onsager-Samaras 
law \cite{Onsager}.  In the thin  interface 
limit $\xi \rightarrow 0$, 
Levin and Flores-Mena \cite{Levin} 
argued that ions in  water cannot approach  the interface  
within the distance of the hydration shell radius 
 $R_{\rm shell}^i$ \cite{Is} (on the order of the  size  
of a water molecule).  In our scheme, 
for finite interface thickness $\xi$, 
hydrophilic   ions are repelled from an interface 
into the $\alpha$ region  due to  the solvation 
interaction (due to the factor $e^{g_i\phi}$ in Eq.(2.9) 
for $g_i>0$). 
Thus,  even in the absence of the image interaction, 
a depletion layer of hydrophilic ions 
can be formed and  the linear behavior 
$\Delta\gamma \propto  n_\alpha$ 
still follows.  To make  qualitative arguments, therefore, 
the image interaction may be neglected for not very large 
$A$ in Eq.(2.6).

\begin{figure}[h]
 \includegraphics[scale=0.52]{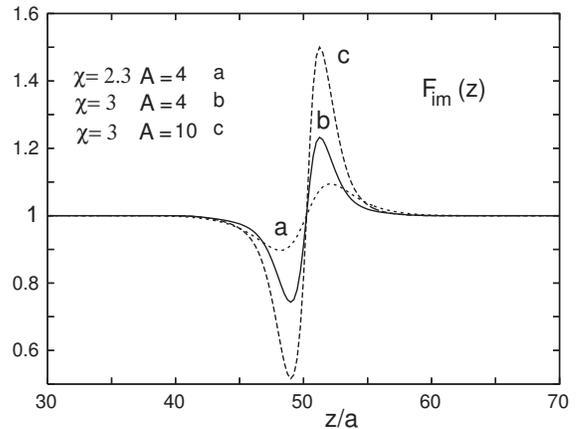} 
\caption{
Image factor $F_{\rm im}(z)$ for monovalent 
ions around an interface 
for (a) $\chi=2.3$ and $A=4$, (b) $\chi=3$ and $A=4$, 
and (c) $\chi=3$ and $A=10$, where $
c_{1\alpha}=c_{1\beta}=2\times 10^{-4}$.
}
\end{figure}

\begin{figure}[h]
\includegraphics[scale=0.52]{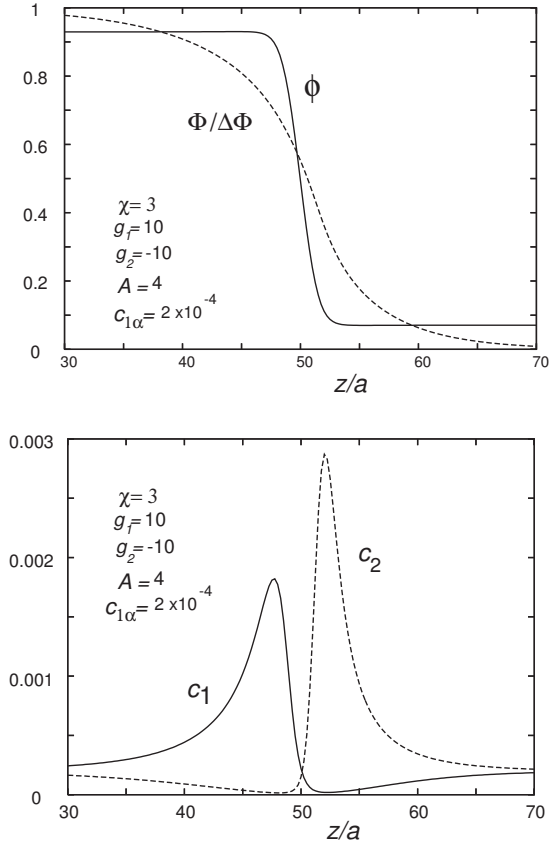}
\caption{Normalized electric potential $\Phi(z)/\Delta\Phi$ 
and composition $\phi(z)$ (upper panel), 
and normalized ion densities $c_1(z)$ and $c_2(z)$ 
(lower panel), where $\chi=3$, $A=4$, 
$g_1=-g_2=10$, and $c_{1\alpha}=c_{1\beta}= 
2\times 10^{-4}$.  For this pair of 
hydrophilic and hydrophobic  
ions, a microphase separation 
forming  a  large electric double layer 
is apparent. 
}
\end{figure}
\begin{figure}[h]
\includegraphics[scale=0.5]{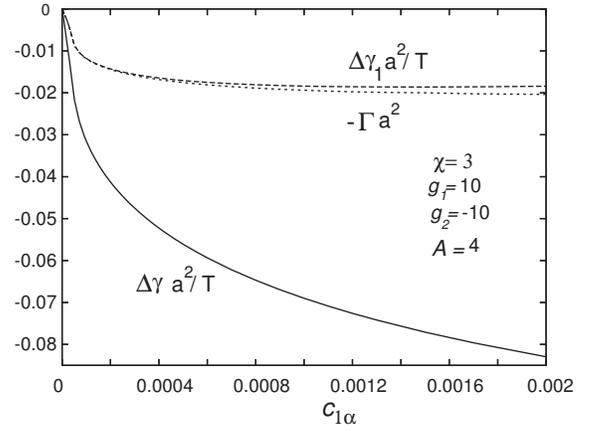}
\caption{   
$a^2\Delta\gamma /T$ 
 and $a^2\Delta\gamma_1 /T$ 
 as functions of $c_{1\alpha}$  for strongly 
hydrophilic and hydrophobic ion pairs, 
where 
$\Delta\gamma=\Delta\gamma_1-|\gamma_{\rm e}|$  and 
$\Delta\gamma_1=\gamma_1-\gamma_{0}$. 
The latter is very close to  
$-a^2\Gamma $ 
in agreement with Eq.(2.22). 
In this case 
$|\gamma_{\rm e}|$ is considerably larger than  
$|\Delta\gamma_{1}|$.  The parameters are the same as in Fig.2.}
\end{figure}
\begin{figure}[h]
 \includegraphics[scale=0.41]{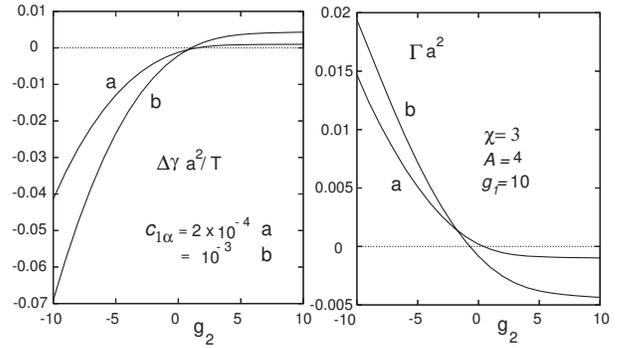}
\caption{ 
$a^2\Delta\gamma/T$ (left) and 
 $a^2\Gamma $ (right) vs $g_2$ 
for $c_{1\alpha}=2\times 10^{-4}$ (a)  
and $10^{-3}$ (b), where $\chi=3$, $A=4$, 
and $g_1=10$. Anions are hydrophilic for $g_2>0$ 
and  hydrophobic  for $g_2<0$. The Gibbs relation 
$\Delta\gamma\cong -T\Gamma$ holds for hydrophilic pairs, 
but does not hold for $-g_2\gg 1$ due to increasing 
$|\gamma_{\rm e}|$. 
}
\end{figure}

In  Fig. 1, 
we show examples of the image factor 
for monovalent ions near an interface, 
\be 
F_{\rm im}(z) = \exp[-\mu_{\rm im}(z)/T], 
\en 
which is calculated from Eq.(2.5) 
for $\ve_1/\ve_c =0.8$ and  $\chi=3$. 
This factor  appears in Eq.(2.9) for $Z_i=\pm 1$. It 
is smaller than unity in the  $\alpha$ region 
and is larger than unity in the  $\beta$ region, 
since the image potential is repulsive  
in the $\alpha$ region  and is attractive in the  
$\beta$ region (see Appendix B of our previous paper \cite{OnukiPRE}). 
At low densities,  
its minimum and maximum are  more enhanced   
for larger  $A\sim \ell_{{\rm B}c}/a$ 
and for larger $\chi$ (larger 
$\Delta\phi$), in accord with Eq.(2.14). See also  
analysis of the image interaction in Figs. 8-10 
of our previous work \cite{OnukiPRE}.

\subsection{Surface tension}

In order to calculate the surface tension, 
 we introduce the grand potential density,
\be 
\omega= f- \sum_i\mu_i n_i- h\phi,
\en  
where $f$ is given by Eq.(2.1). The 
 space integral of $\omega$  is minimized under 
given boundary conditions  in equilibrium. 
In Appendix B, we  will examine how 
$\omega(z)$ is related to  the bulk pressure 
and the surface tension.  We shall see that 
 the discontinuity of  $\omega$ 
across the interface $\omega_\alpha-\omega_\beta$ 
is nearly equal to 
the minus of the pressure discontinuity 
$p_\beta- p_\alpha$  in equilibrium. 
For a planar interface we have $\omega_\alpha
=\omega_\beta=\omega_\infty$.

Using Eq.(2.1) and eliminating 
$\mu_i$ with the aid of Eq.(2.8),  
we obtain 
\be 
\omega= f_0 + \frac{C}{2}|\nabla \phi|^2- h\phi 
 - Tn + \frac{\ve(\phi)}{8\pi}{ E}^2-\rho\Phi
\en 
where $n=\sum_in_i$ is the total ion density. 
Since  $\omega(z)\rightarrow  \omega_\infty$ 
as $z\rightarrow \pm\infty$ for a planar interface, 
 $h$ is expressed as  
\be 
h=[f_0(\phi_\alpha)-f_0(\phi_\beta)]/\Delta\phi 
-T \Delta n /\Delta\phi ,
\en 
where $\Delta  n= n_\alpha-n_\beta$.  
With Eqs.(2.13) and (2.18) 
the bulk volume fractions $\phi_\alpha$ and $\phi_\beta$ 
are determined  for given bulk ion densities (see Appendix A). 
 The surface tension is  expressed as 
$\gamma= \int dz [\omega(z)-\omega_\infty]$ (see Appendix B). 
It  consists of  two contributions 
as $\gamma=\gamma_1+\gamma_{\rm e}$ with   
\bea 
&&\hspace{-1cm}\gamma_1=\int dz [ f_0(\phi) 
 + \frac{C}{2}\phi'^2- h\phi  
 - Tn   -C_\alpha],\\
&&\hspace{-1cm}\gamma_{\rm e}=-\frac{1}{2}\int dz \rho\Phi= 
-\frac{1}{8\pi}
 \int dz{\ve(\phi)}{ E}^2 ,
\ena  
where $\phi'=d\phi/dz$ and 
$C_\alpha= 
f_0(\phi_\alpha)- h\phi_\alpha-Tn_\alpha$. 
From Eq.(2.18)  the  integrand of Eq.(2.19) 
vanishes in the bulk $\alpha$ and $\beta$ regions. 
The lower and upper bounds of the integrations 
in Eqs.(2.19) and (2.20) are pushed to infinity.

Here it is convenient  to introduce 
the excess adsorption  
$\Gamma$ of the ions onto the interface  by
\bea 
\Gamma&=& \int dz[ n-n_\alpha -\frac{\Delta n}{\Delta\phi}
 (\phi-\phi_\alpha)]\nonumber\\
&=& \int_0^{z_{\rm int}}
 dz( n-n_\alpha)+\int_{z_{\rm int}}^L 
 dz( n-n_\beta). 
\ena
In the first line the integrand vanishes 
far from the interface. 
In the second line we suppose a finite system 
in the region $0<z<L$ with $L\gg \xi$ 
and determine the interface position 
 by  
$
 z_{\rm int}= 
\int_0^L dz [\phi(z)-\phi_\beta]/\Delta \phi.
$  
In Appendix A we shall see that 
$\gamma_1$ is related to $\Gamma$ at low ion densities by    
\be 
\Delta\gamma_1= \gamma_1- \gamma_0\cong -T\Gamma,
\en  
where $\gamma_0$ is the surface tension without ions.  
The so-called Gibbs adsorption equation \cite{Widom} relates 
the surface tension and 
the excess adsorption 
by  $(\p \gamma/\p \ln n)_T = -T \Gamma$, where 
$n$ is the number density of the doped particles 
(the ion density in our case). This equation yields Eq.(2.22) 
if $\Gamma \propto n$ and if $\gamma_{\rm e}$ is neglected. 
In our problem, however,  Eq.(2.22) can be derived  
even if $\Gamma$ is not a linear function of $n$ 
(as in Fig.3).

The $\gamma_{\rm e}$ in Eq.(2.20) 
is the minus of the $z$ integration  of the 
electrostatic energy density 
$f_{\rm el}= \ve E^2/8\pi$. 
It arises  from the last two terms in Eq.(2.17)
with the aid of 
$\int dz\rho\Phi= \int dz \ve(\phi){ E}^2/4\pi$. 
In all the examples in 
our previous work \cite{OnukiPRE}, 
it was   at most a few percents 
of $\Delta\gamma=\gamma-\gamma_0$.  
However, in the low density limit, we shall see that 
$\gamma_{\rm e}$ is proportional to the square root of 
the ion density.  Moreover,  its 
magnitude will turn out to be enlarged  for 
strongly hydrophilic and hydrophobic ion pairs. 
This square-root dependence can be 
understood easily as follows.  For simplicity, if 
$|g_1|$ and $|g_2|$ are of order unity, 
the screening lengths in the two phases, 
are  of the same order and $\Phi(z)$ changes 
from $\Phi_\alpha$ to $\Phi_\beta$ smoothly  
on the scale of the screening length 
$\kappa_\alpha^{-1}\sim 
\kappa_\beta^{-1}$. Thus  
$E(z) \sim \Delta\Phi\kappa_\alpha$ 
for $|z|\ls 1/\kappa_\alpha$ 
and  $\gamma_e \sim - (\Delta \Phi)^2
\ve_\alpha\kappa_\alpha/4\pi$.

\section{Two  species of ions} 
\setcounter{equation}{0}

In  the presence of two species of ions, 
where  $Z_1>0$ and $Z_2<0$, we have 
the common density ratio ${n_{1\alpha}}/{n_{1\beta}}
={n_{2\alpha}}/{n_{2\beta}}$. In terms of $g_i\Delta\phi$, 
the ion density ratios and 
the Galvani potential difference are given by  
\bea 
\frac{n_{1\alpha}}{n_{1\beta}} &=& 
\exp\bigg[\frac{|Z_2|g_1+Z_1g_2}{Z_1+|Z_2|}  \Delta\phi\bigg],
\\ 
\frac{e}{T}\Delta\Phi &=&\frac{g_1-g_2}{Z_1+|Z_2|} \Delta \phi . 
\ena  
The Galvani potential difference 
is created by an electric double layer at the interface. 
There is no electric field for the symmetric case $g_1=g_2$ 
in our theory.  The potential $\Phi(z)$ changes 
on the scale of $\kappa_\alpha^{-1}$ 
in the $\alpha$ region and  
on the scale of $\kappa_\beta^{-1}$ 
in the $\beta$ region far from the interface at low ion densities. 
When $\kappa_\beta^{-1}\rightarrow \infty$ 
(as in the water-air case), $\Phi(z)-\Phi_\alpha$ 
becomes very small around the interface 
changing slowly in the $\beta$ region 
\cite{OnukiPRE}, as has been assumed in  the literature 
\cite{Wagner,Onsager,Levin}.

\subsection{Numerical results}

We numerically 
seek equilibrium interface 
solutions of Eqs.(2.8) and (2.10) 
for $C=\chi$ and $\ve_1/\ve_c=0.8$ as in our previous 
work\cite{OnukiPRE}.  Hereafter $c_{1\alpha}=c_{2\alpha}$  
and  $c_{1\beta}=c_{2\beta}$ in the 
monovalent case. We set $\chi=3$ and $A=4$, 
except in Fig.1. We  have 
$\gamma_0= 0.497T/a^2$  without salt and 
$\Delta\phi=0.93$ 
at very small ion densities (see Appendix A)  at $\chi=3$. 
The dielectric constant $\ve_\alpha$ in the $\alpha$ phase 
 is twice larger than that   $\ve_\beta$ 
 in the $\beta$ phase.
We vary the solvation parameters 
$g_i$ and the ion densities in the following figures.

In Fig. 2, we show a set of equilibrium  profiles 
for  $g_1=-g_2=10$ in the monovalent case \cite{comment}. 
Here  $c_{1\alpha}$ and $c_{1\beta}$ coincide 
 from  Eq.(3.1) and is set equal to 
$2\times 10^{-4}$ \cite{cc}.  
The interface thickness 
$\xi$ is of order $5a$, while $\Phi(z)$ changes on the scale of 
$\kappa_\alpha^{-1}\sim \kappa_\beta^{-1}\sim 10a$. 
Here $\Delta\gamma= -0.041Ta^{-2}$ and 
$\Gamma= 0.014a^{-2}$.  We can see a marked growth of  
the electric double layer, which is enhanced 
with increasing $g_1$ for the case $g_1=-g_2$. 
The ion density $c_1+c_2$ has a deep minimum at 
the interface position, for which see Fig. 5 also. 
In our previous work \cite{OnukiPRE}, 
we obtained  milder ion profiles 
for $g_1=-g_2=4$ and $\chi=2.3$.

In Fig. 3, we examine  how 
the surface tension $\gamma$ 
is decreased with 
increasing $c_{1\alpha}(=c_{1\beta})$, where 
$g_1=-g_2=10$ as in Fig. 2. 
We notice the following. (i) The changes  
$\Delta\gamma=\gamma-\gamma_0$ 
and $\Delta\gamma_1=\gamma_1-\gamma_0$ are  
both proportional to  $c_{1\alpha}^{1/2}$ at small 
$c_{1\alpha}$. Here $|\Delta\gamma|/c_{1\alpha}^{1/2}$ 
 is of order unity, so  $\Delta\gamma$    
is appreciable even for very small 
$c_{1\alpha}$.   On the other hand,  for hydrophilic 
ion pairs \cite{OnukiPRE},  
$\Delta\gamma\cong A_\lambda c_{1\alpha}T/a^2>0$ with 
the coefficient  $A_\lambda$ being of order unity,  
which is consistent with the well-known 
 surface tension increase  
for  water-air interfaces with salt. 
(ii) Comparing $\Delta\gamma_1$ 
and $\Delta\gamma=\Delta\gamma_1+\gamma_{\rm e}$, 
we recognize that $\gamma_{\rm e}$ is 
a dominant negative contribution 
in this case. (iii) We confirm 
that  Eq.(2.22) holds excellently (see Appendix A).

In Fig. 4, we 
display $\Delta\gamma$ and $\Gamma$ 
as functions of  $g_2$ in the range $[-10,10]$ 
at $g_1=10$. For $g_2>0$,  
we have the usual behavior  $\Delta\gamma>0$ and $\Gamma<0$.    
 For $g_2<0$,  their signs are reversed  and 
their magnitudes are increased dramatically.

\subsection{Analysis using  the Poisson-Boltzmann equation}

\begin{figure}[h]
 \includegraphics[scale=0.52]{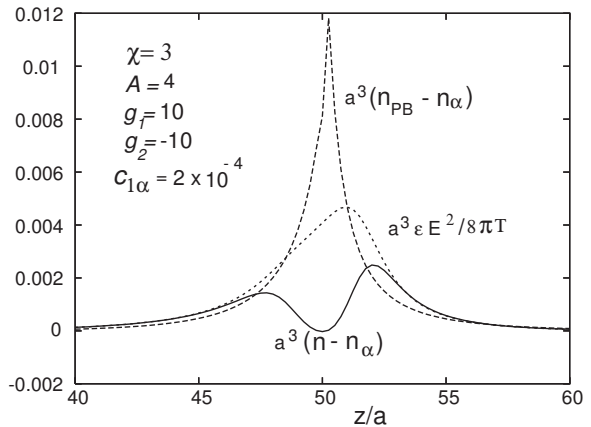}
\caption{Normalized electrostatic 
energy density $a^3 \ve(\phi) E(z)^2/8\pi T$ and 
ion density deviation 
$a^3(n(z)-n_\alpha)$ calculated numerically, which are 
compared with the Poisson-Boltzmann 
solution $a^3(n_{\rm PB}(z)-n_\alpha)$ in Eq.(3.4). 
The $z$ integration and division by 
$-a^3/T$ of the first quantity  
is equal to $\gamma_{\rm e}$ in Eq.(2.20)  and that of 
the third quantity is equal to 
$\gamma_{\rm e}^{\rm PB}$ in Eq.(3.7).   
Here $\chi=3$,  $A=4$, and  $
c_{1\alpha}=c_{1\beta}=2\times 10^{-4}$. }
\end{figure}
\begin{figure}[h]
 \includegraphics[scale=0.5]{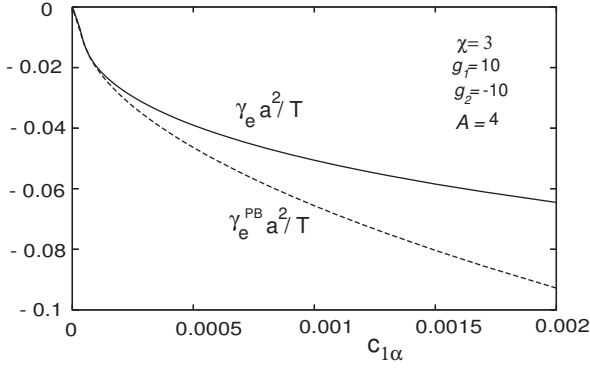}
\caption{
Numerical $a^2\gamma_{\rm e}/T$ in Eq.(2.20) and 
approximate $a^2\gamma_{\rm e}^{\rm PB} /T$ 
in Eq.(3.7) as functions of $c_{1\alpha}$, 
which agree at small $c_{1\alpha} (\ls 10^{-4})$.}
\end{figure}

Since $\gamma_{\rm e}$ is dominant  in Fig. 3, 
we seek its approximate expression to examine its overall behavior. 
To  this end, we consider situations 
in which  the image interaction 
is not crucial,  the ion densities 
are very low, and the ions are monovalent.   Then, 
far from the interface position $|z-z_{\rm int}|>\xi$, 
the ion distributions obey 
the nonlinear Poisson-Boltzmann equation. 
That is, for the normalized potential 
$U(z)=e\Phi(z)/T$,   we assume  
$dU(z)^2/dz^2= \kappa_\beta^2\sinh (U(z)-U_\beta)$ 
in the region $z>z_{\rm int}$ 
and $dU(z)^2/dz^2= \kappa_\alpha^2\sinh(U(z)-U_\alpha)$ 
in the region $z<z_{\rm int}$, 
where $U_K=e\Phi_K/T$  
and $\kappa_K= (4\pi n_Ke^2/\ve_K T)^{1/2}$ 
($=$ Debye wave number) 
in the two phases, $K=\alpha$ and $\beta$. 
Here we are supposing  
 the thin interface limit $\xi\rightarrow 0$.

The resultant electric potential 
 $\Phi_{\rm PB}$ is then given by   
\be
\Phi_{\rm PB}(z) =\Phi_K + 
\frac{2T}{e} \ln \bigg   
[\frac{1+ d_K e^{-\kappa_K|z-z_{\rm int}|}
}{1-d_K e^{-\kappa_K|z-z_{\rm int}|}}\bigg ],  
\en 
where $K=\alpha$  in the region $z<z_{\rm int}$ 
and $K=\beta$  in the region $z>z_{\rm int}$. 
The coefficients  $d_\alpha$ and $d_\beta$ 
are determined from 
 the continuity of $\Phi_{\rm PB}$ and 
$\ve d\Phi_{\rm PB}/dz $ at 
$z=z_{\rm int}$. In 
this approximation,  
the electrostatic energy density divided by $T$ 
and the ion number density coincide  as 
\be
\frac{\ve E_{\rm PB}^2}{8\pi T}
 =n_{\rm PB}(z)-n_K
= \frac{8n_Kd_K^2 e^{-2\kappa_K|z-z_{\rm int}|}}{[1-d_K^2 e^{-2\kappa_K|z-z_{\rm int}|}]^2},
\en  
where $E_{\rm PB}= -d\Phi_{\rm PB}/dz$ is the electric field and 
 $n_{\rm PB}(z)=n_K\cosh [U(z)-U_K]$ 
is the number density with $n_K=n_{1K}+n_{2K}$ 
being the bulk densities.  The potential value 
at the interface position is given by  
\be
\frac{e}{T}[\Phi_\alpha-\Phi(z_{\rm int})] 
= \ln \bigg[
\frac{1+b e^{e\Delta\Phi/2T}}{1+b e^{-e\Delta\Phi/2T}} 
\bigg],
\en
where 
$b= {\ve_\beta\kappa_\beta}/{ \ve_\alpha \kappa_\alpha}
={ n_\beta}{\kappa_\alpha}/{ n_\alpha}{\kappa_\beta}$ 
with $\ve_\alpha$ and $\ve_\beta$ being 
 the dielectric constants  in the bulk phases. In our scheme, 
Eqs.(3.1) and (3.2) give  
$e\Delta\Phi/2T= (g_1-g_2)\Delta\phi/4$ and 
\be 
b= 
({\ve_\beta /\ve_\alpha})^{1/2} 
\exp[- ({g_1+g_2})\Delta\phi/4].
\en  
Note that $\Delta\Phi$ and $b$ are independent of 
the ion density.

In Fig. 5, we compare numerically calculated  
$n-n_\alpha$  and  $\ve E^2/8\pi$ 
with  the Poisson-Boltzmann 
solution ${\ve E_{\rm PB}^2}/{8\pi T}=
n_{\rm PB}- n_\alpha$ in Eq.(3.4). 
Here,  $\chi=3$,  $A=4$, and  $
c_{1\alpha}=c_{1\beta}=2\times 10^{-4}$, for which 
the image interaction is not severe as shown in Fig. 1.  
Far from the interface the Poisson-Boltzmann 
solution is  a good approximation for these two 
quantities, while   at $z\sim z_{\rm int}$ 
it  neglects  the double-layer structure and 
cannot describe the deep minimum of $n-n_\alpha$. 
We notice that the $z$ integration  of 
 ${\ve E_{\rm PB}^2}/{8\pi T}
=n_{\rm PB}- n_\alpha$ is  close to that 
of $\ve E^2/8\pi$  but is a few times larger than that of 
$\Gamma$ in this case. 
Hence we approximate $\gamma_{\rm e}$ in  Eq.(2.20) as the integral of 
$-{\ve E_{\rm PB}^2}/{8\pi T}= 
-n_{\rm PB}+n_\alpha$, denoted by   
$\gamma_{\rm e}^{\rm PB}$. 
Some calculations yield 
\bea
\frac{\gamma_{\rm e}^{\rm \small{PB}}}{T} &=&  
\frac{2n_\alpha}{\kappa_{\alpha}}\bigg 
[1+b- \sqrt{1+b^2+
2b\cosh\bigg(\frac{e\Delta \Phi}{2T}\bigg)}\bigg] \nonumber\\
&=& - A_s (n_\alpha/\ell_{{\rm B}\alpha} )^{1/2} , 
\ena 
where   $b$ is defined by Eq.(3.6) and  
$\ell_{{\rm B}\alpha}=e^2/\ve_\alpha T$ 
 is the Bjerrum length 
in the $\alpha$ phase. In the second line, the 
dimensionless coefficient 
$A_s$  is determined by  
$b$ and $e\Delta\Phi/T$ and is independent of 
the ion density $n_\alpha$, so  
$\gamma_{\rm e}^{\rm PB}\propto -n_{\alpha}^{1/2}$. 
Thus  we obtain the result 
$\Delta\gamma\propto -{n_\alpha}^{1/2}$  
at low ion densities for  asymmetric  ion pairs $g_1\neq g_2$. 
For small $e\Delta\Phi/T$, 
we have $\gamma_{\rm e} \sim -(\Delta\Phi)^2\ve_\alpha
\kappa_\alpha/4\pi$ 
in accord  with the argument at the end of Section 2. 
Also the excess adsorption $\Gamma$ 
should exhibit this ion-density dependence, 
since   the long-range tail of 
$n-n_\alpha$ and that of $n_{\rm PB}-n_\alpha$ 
should coincide.

In the case $g_1\ge -g_2 \gg 1$, 
we have $e\Delta \Phi/T = (g_1+|g_2|)\Delta\phi/2$ 
and $b\ll  e^{e\Delta \Phi/2T}$ in Eq.(3.7).  Then,  
\be 
A_s  \cong  
 \pi^{-1/2} (\ve_\beta/\ve_\alpha)^{1/4}
e^{|g_2|\Delta \phi/4}, 
\en 
which grows 
with increasing $g_1$ and $|g_2|$.   
 In Fig. 3, we notice that  the coefficient 
in front of $c_{1\alpha}^{1/2}$ 
in $\Delta\gamma a^2/T$ can be of 
order  unity, which is equal to $A_s(\pi/4A)^{1/2}$ from Eq.(3.7). 
In Fig. 6, we compare this  approximate 
$\Delta\gamma_{\rm e}^{\rm PB}$ 
and the numerical $\Delta\gamma_{\rm e}$ 
for $g_1=-g_2=  10$ as functions of 
$c_{1\alpha}$.  Agreement is 
 excellent for $c_{1\alpha}\ls 10^{-4}$, while 
$|\Delta\gamma_{\rm e}^{\rm PB}|$ is 
larger than  $|\Delta\gamma_{\rm e}|$ for larger $c_{1\alpha}$.

\subsection{Hydrophilic ion pairs}

On the basis of the Poisson-Boltzmann 
theory, we may also examine  the usual 
case of hydrophilic ion pairs, 
where  $g_1$ and $g_2$ are both considerably 
larger than unity with $g_1>g_2$.  
(Here $\Delta\gamma_{\rm e}^{\rm PB} =0$ 
 for $g_1=g_2$.) Then  
we have $n_\alpha\gg n_\beta$ and  
$b \ll 1$   not close to 
the critical point, leading to 
\be 
\Delta\gamma_{\rm e}^{\rm PB}  \cong   
-2T[\cosh(e\Delta\Phi/2T) -1] 
b{n_\alpha}/{\kappa_\alpha} ,
\en   
where $b{n_\alpha}/{\kappa_\alpha}= 
{n_\beta}/{\kappa_\beta}\propto {n_\beta}^{1/2}$.
Here the right hand side is the integration result 
in the $\beta$ region, since 
that in the $\alpha$ region is smaller by 
$b$.  From Eq.(3.5) $\Phi_\alpha -\Phi(z) 
\propto b$ in the $\alpha$ region. 
If $(g_1-g_2)\Delta\phi>1$, 
we find $|\Delta\gamma_{\rm e}^{\rm PB}|a^2/T\sim 
e^{-g_2\Delta \phi/2}(c_{1\alpha}/A)^{1/2}$, where 
$A$ is defined by  Eq.(2.6).  Thus, together with positive 
$\Delta\gamma_1$, we obtain the following expression, 
\be 
\Delta \gamma \cong - A_sT 
(n_\alpha/\ell_{{\rm B}\alpha} )^{1/2} 
+ \lambda_s T n_\alpha.   
\en 
The coefficient $A_s$ 
is  small here ($\sim (\ve_\beta/\ve_\alpha)^{1/2}
e^{-g_2\Delta \phi/2}$). 
The second term 
is of the  well-known form accounting for the  ion depletion 
near the interface.  
With this expression,  
 $\Delta\gamma$ should exhibit a small 
minimum given by  
\be 
(\Delta\gamma)_{\rm min}
=-T\lambda_s n_{\rm m}
\en 
at  $n=n_{\rm m}= (A_s/2\lambda_s)^2/\ell_{{\rm B}\alpha}$. 
As an example, 
let the ion concentration 
giving this minimum be $1$ mM  in 
the water-rich phase  \cite{cc}. 
Then we obtain  $A_s= 1.2\times 10^{-2}$ 
by setting $\lambda_s=3{\rm  \AA}$ and 
$\ell_{{\rm B}\alpha} =7{\rm \AA}$.

For  water-air interfaces, 
Jones and Ray \cite{Jones}
found a negative minimum in $\Delta\gamma$ 
of order $-10^{-4}\gamma_0$. 
Their data can  well be 
fitted to Eq.(3.10) with $A_s \sim 10^{-2}$. 
However, we have assumed 
appreciable ion densities even 
in the less polar $\beta$ region. 
That is, our one-dimensional 
 calculations  are justified only when 
the screening length 
$\kappa_\beta^{-1}$ in the $\beta$ region 
is much shorter than any characteristic lengths 
in experiments, which  are the inverse curvature 
of the meniscus or the wavelength 
of capillary waves, for example. 
In the literature \cite{Wagner,Onsager,Levin,Ka,Manciu} 
ions are treated to be  nonexistent 
in the air region, so in our scheme we do not still 
 understand the Jones-Ray effect.

Here it is worth noting that 
Nichols and Pratt \cite{Pratt} theoretically derived  
the  square root dependence of $\Delta\gamma$ 
as the low-density asymptotic law 
when ions are appreciably soluble  
both in the two phases. 
They calculated  the ion density deviations 
decaying on the scale of the screening length 
far from the interface. which vanish as $\kappa_\beta
\rightarrow 0$ as in our theory.

\section{Three species of ions}
\setcounter{equation}{0}

\begin{figure}[h]
\includegraphics[scale=0.39]{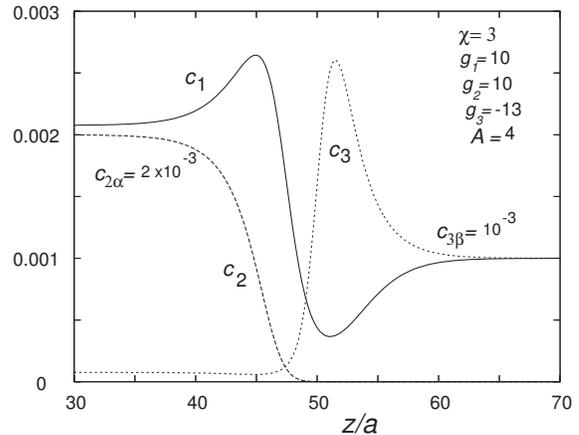}
\caption{
Normalized ion densities 
$c_1(z)$, $c_2(z)$,  and $c_3(z)$ of 
 three ion species  in the monovalent case 
for $c_{2\alpha}=2\times 
10^{-3}$ and $c_{3\beta}=10^{-3}$. 
The  species 1 and 2 are hydrophilic as 
 $g_1=g_2=10$, while  the  third species 
is hydrophobic as $g_3=-13$. 
Then $c_{2\beta}\cong 0$ and 
$c_{3\alpha}\ll 1$. 
}
\end{figure}
\begin{figure}[h]
 \includegraphics[scale=0.36]{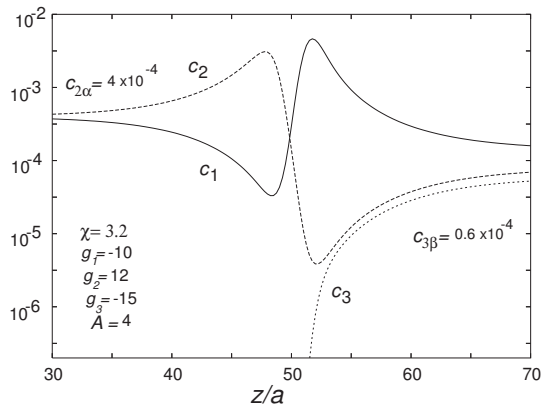}
\caption{
Normalized ion densities 
$c_1(z)$, $c_2(z)$,  and $c_3(z)$ 
 in the presence of 
 three ion species   
with   $g_1=-10$, $g_2=12$,
 and $g_3=-13$ on a semi-logarithmic scale, 
resembling to those in the experiment \cite{Luo}. 
The third species  does not penetrate 
into  the $\alpha$ region. 
}
\end{figure}

Next we consider the ion distributions 
in the case of three ion species  for water-oil interfaces. 
As in the experiment \cite{Luo}, 
 we assume  monovalent 
ions   with   charges 
 $Q_1= e, Q_2= -e,$ and $Q_3= -e$ for the three species. 
Namely, the first species consists of  monovalent cations,  
while the second and third species are monovalent anions. 
The ion distributions are  very complicated and 
 numerical calculations are needed for 
their analysis.

Before presenting numerical results, we first examine 
the potential difference $\Delta\Phi$,  
which is determined from Eqs.(2.12) and (2.13) \cite{Nitro,Hung}. 
For example, in  terms of the ion  densities 
in the  $\alpha$ phase, it follows 
the equation for $\Delta \Phi$ in the form,  
\be
c_{1\alpha}e^{2e\Delta \Phi/T}
=c_{2\alpha}e^{(g_1-g_2)\Delta\phi}+
c_{3\alpha}e^{(g_1-g_3)\Delta\phi},
\en 
where $c_{1\alpha}=c_{2\alpha}+c_{3\alpha}$. 
This yields  Eq.(3.2) for  $c_{3\alpha}=0$ 
in the monovalent case.

However, in the experiment \cite{Luo}, 
the third species (TPB$^-$) was  strongly hydrophobic 
such that $c_{3\alpha}\ll c_{3\beta}$ was realized, while 
the second species (Br$^-$) was hydrophilic.  
Supposing  such cases, let us assume 
$g_2>0$ and $g_3<0$ and 
choose $c_{2\alpha}$ and $c_{3\beta}$ as 
control parameters. If we set  
\be 
X= \exp[e\Delta \Phi/T- (g_1-g_2)\Delta\phi/2],
\en 
Eq.(4.1) becomes a cubic equation, 
\be 
X-X^{-1}=2R 
[1- X^2e^{(g_3-g_2)\Delta\phi}] . 
\en  
where $R(\propto c_{3\beta})$ is  defined by  
\be 
R=  e^{(g_1+g_2)\Delta\phi/2}{c_{3\beta}}/2{c_{2\alpha}}.
\en 
The right hand side of Eq.(4.3) arises 
in the presence of the third species. 
We may well assume 
$X^2e^{(g_3-g_2)\Delta\phi}\ll 1$ 
for large $|g_3| \gg 1$ to obtain 
$
X\cong R+ \sqrt{1+R^2}$ or 
\be 
\frac{e}{T} \Delta \Phi \cong \frac{1}{2}(g_1-g_2)\Delta\phi+ 
\ln(R+ \sqrt{1+R^2}).
\en 
If $(g_1+g_2)\Delta\phi \gg 1$, we readily reach  
the new regime $R \gg 1$  even for  small 
$c_{3\beta}$, where $X\cong 2R$ holds 
and $X^2e^{(g_3-g_2)\Delta\phi}
\cong ({c_{3\beta}}/{c_{2\alpha}})^2
e^{(g_1+g_3)\Delta\phi}$ needs to be small. That is,  
we find 
\be 
\frac{e}{T}\Delta \Phi  \cong 
g_1\Delta\phi    
+ \ln ({c_{3\beta}}/{c_{2\alpha}}) 
\en  
for   
$ e^{-(g_1+g_2)\Delta\phi/2} \ll c_{3\beta}/c_{2\alpha} 
\ll  e^{-(g_1+g_3)\Delta\phi/2}$. 
In this case, 
$c_{1\alpha}\cong c_{2\alpha}$, $c_{1\beta}\cong c_{3\beta}$, 
$c_{2\beta}\cong  e^{-(g_1+g_2)\Delta\phi}
c_{2\alpha}^2/{c_{3\beta}}\ll c_{3\beta}$,  
and 
$c_{3\alpha}\cong 
 e^{(g_1+g_3)\Delta\phi} c_{3\beta}^2/c_{2\alpha}\ll 
c_{2\alpha}$.

In Fig. 7, we display the ion distributions 
in the presence  of three ion species in the monovalent case 
with $Q_1=e$, $Q_2=-e$, and $Q_3=-e$. Since the 
 absolute values  of $g_i$ are taken to be large,  
we can see steep and complex  variations of the ion 
distributions around the interface. The first 
 and  second species  are both hydrophilic   but 
the third one  is hydrophobic as $g_1=g_2=10$ and $g_3=-13$. 
Here   $\chi=3$, $e\Delta\Phi/T=7.92$, 
and  $\gamma=0.446T/a^2$, while $\Delta\Phi=0$ 
for $c_{3\beta}=0$ since $g_1=g_2$.  
The peaks of $c_1$ and $c_3$ are conspicuous. 
This is the case discussed around Eq.(4.6), since 
 $X^2e^{(g_3-g_2)\Delta\phi}
 =1.5\times 10^{-2}$ and  $R= 2.7\times 10^3$.

In Fig. 8,  the first  and third 
species are hydrophobic but  the second 
species is hydrophilic as  $g_1=-10$, $g_2=12$, and $g_3=-15$, 
where $\chi=3.2$, $e\Delta\Phi/T=7.01$, and   
 $\gamma=0.600T/a^2$ (with $\gamma_0=0.620T/a^2$ at  $\chi=3.2$). 
Here  $c_1$ and $c_2$ exhibit 
sharp peaks. The ion distributions in this  
case can  be compared with   those  in 
 the experiment by Luo {\it et al.}\cite{Luo}, so  
 the curves  are written on 
a semilogarithmic scale as in their paper. 
The adopted parameter values    are 
inferred from their  experimental data.

\section{Summary}

Effects of ions  in polar fluid mixtures 
are very complex because of the presence of 
the electrostatic, solvation, and image interactions. 
Our continuum theory takes  account 
of these interactions, 
though it  should be inaccurate 
on the angstrom scale. 
Our main results are as follows. 
In Fig. 2,  we have illustrated 
the singular ion distributions 
around an interface  
when hydrophilic and 
hydrophobic ions coexist.   
We have given the general expression for the 
surface tension in electrolytes in Eqs.(2.19) and (2.20). 
In Fig. 3, the resultant surface tension change  $\Delta\gamma$  
 is proportional to the square root of 
the ion density in the dilute limit.   This dependence arises from the 
fact that the electrostatic field 
changes over the distance of the screening length far 
from the interface. 
When the image 
interaction is not severe, 
the electrostatic contribution 
$\gamma_{\rm e}$ to the surface tension 
 can be estimated 
by the Poisson-Boltzmann result in Eq.(3.7), 
as demonstrated in Fig. 6. 
For hydrophilic ion pairs, 
we propose the expression (3.10) 
consisting of the negative square-root and positive 
linear terms.  In the presence of three ion species, 
the ion distributions are very complex as in Figs. 7 and 8, 
where Fig. 8 corresponds to the experimental result \cite{Luo}.

Using salts composed of 
hydrophilic and hydrophobic ions,  
 a large decrease of $\gamma$  far from the 
critical point \cite{Nitro,Luo}, 
ion distributions near an interface \cite{Luo},   and 
a mesophase near the critical point \cite{Seto} 
have already been reported. 
We propose more systematic surface-tension measurements 
of water-oil interfaces  with  dissimilar 
cations and anions.

In this paper, we have neglected the  
clustering of ions.  (i) For hydrophilic ions, 
this effect becomes conspicuous for densities 
 larger than 1M   in aqueous solutions 
\cite{pair,pair1,pair2}. 
It is not clear how hydrophilic and hydrophobic ions  
can form clusters with increasing their densities. 
(ii) It is worth noting that  ion aggregation 
was predicted to occur near  the critical point 
(even among ions of the same species) 
because of a long-range   interaction 
mediated by the critical fluctuations 
(see Appendix A of our previous 
paper \cite{OnukiPRE}). 
(iii) For ionic surfactants, 
the amphiphilic interaction with water and oil  needs to be 
included. Recently we have found that 
the adsorption of ionic surfactant and counterions 
onto a water-oil  interface 
is dramatically  enhanced when these two species constitute 
a hydrophilic and hydrophobic pair 
producing  a large  electric double 
layer  \cite{OnukiEPL}.

In future,   
we should study 
the surface tension near the critical point 
in the presence of hydrophilic and hydrophobic ions with $|g_i|\gg 1$. 
The  mesophase formation realized 
for such ion pairs  near the critical point  
should also be investigated  
\cite{Onuki-Kitamura,OnukiPRE,Seto}. 
Dynamics of phase ordering 
in  ionic fluids  with the solvation 
interaction has  
not yet been explored.  
Transient relaxations  under applied electric field 
could   also be  studied.

\acknowledgments
{This work was  
 supported by   Grants in Aid for Scientific 
Research and for  
the 21st Century COE project (Center for Diversity and Universality in
Physics) from the Ministry of Education, Culture, Sports, Science and 
Technology of Japan.}

\vspace{2mm} 
{\bf Appendix  A: Derivation of the Gibbs relation 
for  dilute ion densities}\\
\setcounter{equation}{0}
\renewcommand{\theequation}{A\arabic{equation}}

We here derive Eq.(2.22)  in the Ginzburg-Landau 
theory for small $n$. 
Let  $\phi_0(z)$ and $h_0$ be the composition profile 
and the chemical potential difference without ions. From Eq.(2.10) 
the deviation $\delta\phi=\phi-\phi_0$ obeys 
\be 
[f_0''(\phi) -C\frac{d^2}{dz^2} ]\delta\phi
 =  \frac{\ve_1 { E}^2}{8\pi}  +T \sum_i g_in_i 
  - h_{\rm im}+\delta h,
\en 
to linear order, where  $f_0''=\p^2 f_0/\p\phi^2$. 
For the free energy density in Eq.(2.2) we have 
$f_0''= [\phi^{-1}(1-\phi)^{-1}-2\chi]T/a^3$. 
Obviously, the left hand side of Eq.(A1) vanishes 
to linear order if we multiply it  by $\phi'=d\phi/dz$ 
and integrate over $z$. We can   show that 
the right hand side also vanishes 
in the same procedure from 
$\int dz (\ve_1 E^2\phi'/8\pi+\Phi'\rho)=0$ and 
$\int dz (h_{\rm im}\phi'+ \sum_i  \mu_{\rm im}Z_i^2n_i') =0$ 
where the latter follows 
from Eq.(4.24) of our previous paper \cite{OnukiPRE} 
and  the primes denote operating $d/dz$. 
This  orthogonality to $\phi'$ is the solubility 
condition of Eq.(A1) \cite{OnukiPRE}. 
From Eqs.(2.13) and (2.18)  
the deviation $\delta h=h-h_0$ is 
calculated as 
\be
\delta h =-T {\Delta n}/{\Delta \phi}+\cdots.
\en 
The bulk volume fractions are expanded as 
\be  
\phi_K =\phi_{K0} + {T}
\bigg(\sum_ig_in_{iK}
-\frac{\Delta n}{\Delta\phi}\bigg)/f_0''(\phi_K)+\cdots,
\en 
where  $K=\alpha$ and $\beta$. 
The $\phi_{K0}$ are the bulk values without ions. 
From Eq.(2.19) 
we may  expand  $\gamma_1$  with respect to 
 $\delta\phi=\phi-\phi_0$ up to second order as 
\bea 
\gamma_1&=&\gamma_0-T\Gamma +\int dz \frac{C}{2}
(\delta\phi')^2 
\nonumber\\ 
&&\hspace{-1cm}
+ \int dz\bigg[\zeta(z) -\zeta_\alpha  - 
\frac{\Delta\zeta}{\Delta\phi}(\phi(z) -\phi_\alpha) \bigg] 
+\cdots, 
\ena 
where 
$\zeta=f_0''(\phi)   (\delta\phi)^2/2$ 
and 
$\Delta\zeta= \zeta_\alpha-\zeta_\beta$. The terms linear in 
$\delta\phi$ vanish from the interface equation 
$h_0= f_0'(\psi_0)-C\phi_0''$ without ions. 
The third and fourth terms are  of second order in $\delta\phi$ 
and, if they are  neglected, we obtain Eq.(2.22).  
However, the corrections grow near the critical point.

\vspace{2mm} 
{\bf Appendix  B: Grand potential 
and surface tension  in incompressible fluid 
mixtures}\\
\setcounter{equation}{0}
\renewcommand{\theequation}{B\arabic{equation}}

In the text, 
we have  neglected  the deviation of the 
number density $n_t=n_A+n_B$ of the solvent 
from a reference density  $n_t^0$ assuming 
the common molecular size $(a_A=a_B=a)$.  
More generally, 
the system is characterized by the solvent  densities, 
\be 
n_A=n_t\phi,\quad n_B=n_t (1-\phi),
\en 
in addition to  the ion densities  
$n_1$ and $n_2$. We assume 
$\phi=n_A/(n_A+n_B)$, so $\phi$  is  also the molar fraction. 
The density deviation $\delta n_t=n_t-n_t^0$ is  small.  
Hereafter  we suppress the dependence on 
$T$.

For nearly incompressible fluid mixtures, 
the dependence on 
$\delta n_t$ 
may be accounted for if we replace 
$f_0(\phi)$ in Eq.(2.1) by the Helmholtz free energy density, 
\be 
\hat{f}_0(\phi,n_t)
=f_0(\phi)
+ a_0+ a_1(\phi)\delta n_t+ \frac{a_2(\phi)}{2} 
(\delta n_t)^2,
\en  
where $a_0$ is 
 independent of the densities (being dependent only on $T$), 
but $a_1$ and $a_2$ depend on $\phi$ (and  $T$). 
Far from the interface, where the image interaction 
vanishes and the fluid is homogeneous, 
the pressure $p$ is given by 
\be 
p= -f_0-a_0+ a_1n_t^0 +a_2\delta n_t +T \sum_i n_i, 
\en  
up to first order in $\delta n_t$. The last term is the 
contribution from   ions. 
The coefficient $a_2$ is inversely proportional to the 
compressibility $(\p n_t/\p p)_{T\phi}/n_t$ 
and is  large such that $a_2\delta n_t$ is appreciable,
while   $a_2(\delta n_t)^2/2$ is small in Eq.(B2).   
The chemical potentials 
of the two components  $\mu_A=\delta  F/\delta n_A$ and 
 $\mu_B=\delta F/\delta n_B$ are  expressed as 
\bea 
\mu_A&=& (1-\phi)(h+\Delta h)/n_t+ a_1+ a_2\delta n_t, \nonumber\\ 
\mu_B&=& -\phi (h+\Delta h)/n_t+ a_1+ a_2\delta n_t , 
\ena 
where $h$ is given in Eq.(2.10)  and  
\be 
\Delta h= \frac{\p a_1}{\p \phi} \delta n_t
+ \frac{1}{2} \frac{\p a_2}{\p \phi} (\delta n_t)^2.
\en 
We find $h+ \Delta h= n_t(\mu_A-\mu_B)$. 
The expressions for the ion chemical potentials 
$\mu_1$ and $\mu_2$ in Eq.(2.8) are unchanged. 
For nonvanishing  $\delta n_t$ 
we define  the generalized grand potential 
density  by 
\bea 
\hat{\omega}&=&
\hat{f}_0 - n_A\mu_A-n_B\mu_B-n_1\mu_1-n_2\mu_2\nonumber\\
&=& \omega -n_t  \mu_B
 - \phi  \Delta h  
+  (\hat{f}_0-f_0),
\ena 
where $\omega$ is the grand potential density in Eq.(2.17) 
in the incompressible case  
and use has been made of $n_A\mu_A+n_B\mu_B
= n_t\mu_B + \phi(h+\Delta h)$. 
In the bulk regions 
$\hat{\omega}$ should be 
equal to  the  minus of the pressure $p$.  We assume that the electric field 
vanishes far from the interface and 
there is no contribution of the Maxwell stress tensor.

In particular, when the fluid is separated by a planar interface, 
the four chemical potentials $\mu_A, \mu_B,
\mu_1$, and $\mu_2$ become    homogeneous. 
Here $\hat\omega$ tends to a constant $-p_\infty$ 
as $z \rightarrow \pm \infty$. 
In the right hand side of Eq.(B6), 
we  may replace $n_t$ by $n_t^0$  and 
$\hat{f}_0-f_0$  by $a_0$ for small $\delta n_t$. 
Then,  as assumed in the text,   $\omega$ tends to 
$\omega_\infty= 
-p_\infty+ n_t^0\mu_B-a_0$ as $z\rightarrow \pm \infty$   and 
the surface tension is expressed as 
\bea 
\gamma&=&\int dz[\hat{\omega}(z)+ p_\infty] \nonumber\\ 
&\cong& \int dz[\omega(z)-\omega_\infty],
\ena 
where the first line is general and 
 the second line is  its incompressible  limit 
$\delta n_t\rightarrow 0$.

We also consider an equilibrium spherical droplet 
with radius $R$ in one of the two phases, say, the $\alpha$ phase.  
In this case, the total Helmholtz free energy 
$F$ is minimized for given system 
volume $V$ and the total particle numbers, 
$N_A$, $N_B$, and $N_1=N_2$. 
The densities inside and 
outside the droplet are then determined 
 such that the four chemical potentials become  
homogeneous 
(see Chapter 9.1 of the book by the present author \cite{Onukibook}). 
We furthermore  minimize $F$ with respect to 
$R$ in the presence of the surface free energy 
$4\pi \gamma R^2$.  
This yields the 
famous pressure difference 
 $\Delta p= p_\alpha-p_\beta=2\gamma/R$ 
(the Laplace law).  
Thus there arises a difference in 
the bulk values of $\omega$ given by 
\be 
-\omega_\alpha+\omega_\beta=\Delta p= 2\gamma/R.
\en  
which holds in the   limit 
$\delta n_t\rightarrow 0$. 
As  $R\rightarrow \infty$, we have 
$\omega_\alpha=\omega_\beta=\omega_\infty$.



\begin{references}


\bibitem{Wagner} 
C. Wagner, Phys. Z. {\bf 25}, 474 (1924).
\bibitem{Onsager} 
L. Onsager and N. N. T. Samaras, J. Chem. Phys. {\bf 2}, 528 (1934).

\bibitem{Levin}  
Y. Levin and J. E. Flores-Mena, Europhys. Lett. {\bf 56}, 187 (2001).  


\bibitem{Is} J. N. Israelachvili,  
{\it Intermolecular and Surface 
Forces} (Academic Press, London, 1991). 


\bibitem{Jones}  
G. Jones and W. A. Ray, J. Am. Chem. Soc. {\bf 59}, 187 (1937); 
ibid. {\bf 63}, 288 (1941);    
ibid. {\bf 63}, 3262 (1941).   



\bibitem{tension1} 
N. Matubayasi, H. Matsuo, K. Yamamoto, S. Yamaguchi, and A. Matuzawa, J. Colloid Interface Sci. {\bf 209}, 398 (1999). 


\bibitem{Ohshima} H. Ohshima and H. Matsubara, 
Colloid Polym. Sci. {\bf 282}, 1044 (2002). 

\bibitem{Petersen}  
P. B. Petersen and R. J. Saykally, 
J. Am. Chem. Soc. {\bf 127}, 15446 (2005).



\bibitem{Ka}  K. A. Karraker  and C. J. Radke, 
Adv. Colloid Interface Sci. {\bf 96}, 231 (2002).


\bibitem{Manciu}  M. Manciu and E. Ruckenstein, 
Adv. Colloid Interface Sci. {\bf 105}, 10468 (2003).






\bibitem{Nitro} J. D. Reid, O. R. Melroy, and R.P. Buck, 
J. Electroanal. Chem. {\bf 147}, 71 (1983). 
\bibitem{Onuki-Kitamura} A. Onuki and H. Kitamura, 
  J. Chem. Phys. {\bf 121}, 3143 (2004).

\bibitem{OnukiPRE} A. Onuki, Phys. Rev. E {\bf 73}, 021506 (2006). 



\bibitem{Seto} K. Sadakane, H. Seto, H. Endo, and M. Shibayama, 
J. Phys. Soc. Jpn., {\bf 76}, 113602 (2007). 



\bibitem{Luo} 
G. Luo, S. Malkova, J. Yoon, D. G. Schultz, 
B. Lin, M. Meron, I. Benjamin, P. Vanysek, 
 and M. L. Schlossman,  Science, {\bf 311},  216 (2006). 


\bibitem{Benjamin} I. 
Benjamin, Chemical Reviews {\bf 106}, 1212 (2006).
\bibitem{Tobias}  
P. Jungwirth and D.J. Tobias, Chemical Reviews  {\bf 106}, 1259 (2006).

 

\bibitem{Gibbs} J.W. Gibbs, 
  Collected works, vol.1,pp.219-331 (1957), 
New Haven, CT: Yale University Press. 



\bibitem{Widom}
J.S. Rowlinson and B. Widom,
{\it Molecular Theory of Capillarity} 
(Clarendon, Oxford,1989).



\bibitem{pair} L. Degr$\rm\grave{e}$ve and F.M. Mazz$\rm{\acute{e}}$, 
Molecular Phys. 
{\bf 101}, 1443 (2003). 
 
\bibitem{pair1} A.A. Chen and R.V. Pappu, 
J. Phys. Chem. B {\bf 111}, 6469 (2007). 

\bibitem{pair2} S.A. Hassan,  
Phys. Rev. E {\bf 77}, 031501 (2008). 



\bibitem{Onukibook} A. Onuki, {\it Phase Transition Dynamics} 
(Cambridge University Press, Cambridge, 2002).

\bibitem{Safran} 
S.A. Safran, {\it Statistical Thermodynamics of Surfaces, 
Interfaces, and Membranes} (Westview Press, 2003). 


\bibitem{Debye-Kleboth} P. Debye and K. Kleboth, 
  J. Chem. Phys. {\bf 42}, 3155 (1965).


\bibitem{Gros} M. Gros,  S. Gromb, and 
C. Gavach, J. Electroanal. Chem. 
{\bf 89}, 29 (1978).
 
\bibitem{Koryta} J. Koryta, 
Electrochim. Acta, {\bf 24}, 293 (1979); 
ibid. {\bf 29}, 445 (1984).


\bibitem{Hung}  
Le Quoc Hung, J. Electroanal. Chem. 
{\bf 115}, 159 (1980) 
; ibid. {\bf 149}, 1 (1983).

\bibitem{Osakai} T. Osakai and K. Ebina, 
J. Phys. Chem. B {\bf 102}, 5691 (1998). 

\bibitem{estimate}
The Gibbs transfer energy 
 $\Delta G_{\alpha\beta}^{i}$ per mole 
is $34.2$  for Na$^+$, 
 $67.3$ for Ca$^{2+}$, and 
$-35.9$ for BPh$_4^-$ in units of kJ 
for strongly segregated 
water-nitrobenzene  at room temperatures \cite{Hung}. 
Dividing them  by the Avogadro number gives   
$\Delta\mu_{\alpha\beta}^{i}$ per particle.



\bibitem{cc} The density of 1 mM salt is 
$n=6\times 10^{17}$cm$^{-3}$. If $a=3{\rm \AA}$, 
the normalized density 
$c=a^3n$ is equal to $1.6\times 10^{-5}$. 

\bibitem{comment} A 
mesoscopic phase is realized 
if a parameter  $\gamma_{\rm p}$ 
representing the asymmetry of the ion solvation 
exceeds unity  \cite{Onuki-Kitamura,OnukiPRE}. 
For $C=T\chi/a^2$, 
it is equal to $|g_1-g_2|/8\sqrt{\chi A}$ 
in the monovalent case. In particular, we 
have  $\gamma_{\rm p}=5/4\sqrt{3}<1$ 
for the parameter values in Fig. 2, 
$g_1=-g_2=10$, $\chi=3$, and $ A=4$. 
  

  

\bibitem{Pratt} A. L. Nicols and L.R. Pratt, 
  J. Chem. Phys. {\bf 80}, 6225 (1984).

\bibitem{OnukiEPL} A. Onuki, Europhys. Lett. (in press).




 






\end{references}
\end{document}